\documentclass[aps,prl,twocolumn]{revtex4-2}

\usepackage[utf8]{inputenc}
\usepackage{amsmath}
\usepackage{stmaryrd}
\usepackage{color}
\usepackage{mathrsfs}
\usepackage{yfonts}
\usepackage{rotating}
\usepackage{graphicx}
\usepackage{amsfonts}
\usepackage{mathrsfs}
\usepackage{amsmath}
\usepackage[american]{babel}
\usepackage{wasysym}
\usepackage{slashed}
\usepackage{pgfplots}
\usepackage{braket}
\usepackage{graphicx}
\usepackage[caption=false]{subfig}
\usepackage{dsfont}
\usepackage{amsthm}
\usepackage{youngtab}
\usepackage{gensymb}

\usepackage{natbib}
\PassOptionsToPackage{breaklinks}{hyperref}
\usepackage[colorlinks,allcolors=black]{hyperref}

\theoremstyle{definition}

\usepackage{cleveref}
\crefname{proposition}{Proposition}{Propositions}
\crefname{theorem}{Theorem}{Theorems}
\crefname{definition}{Definition}{Definitions}
\crefname{lemma}{Lemma}{Lemmas}
\crefname{figure}{Figure}{Figures}
\crefname{corollary}{Corollary}{Corollary}
\crefname{conjecture}{Conjecture}{Conjectures}
\crefname{section}{Section}{Sections}
\crefname{appendix}{Appendix}{Appendixes}
\crefname{observation}{Observation}{Observation}
\crefname{remark}{Remark}{Remark}
\crefname{example}{Example}{Examples}
\crefname{equation}{Eq.}{Eqs.}
\crefname{table}{Table}{Tables}

\usepackage{tikz}

\usepackage{amssymb}
   \makeatletter
     \renewcommand\@make@capt@title[2]{%
      \@ifx@empty\float@link{\@firstofone}{\expandafter\href\expandafter{\float@link}}%
       {\textbf{#1}}\@caption@fignum@sep#2\quad}%
     \makeatother
\makeatletter
\renewcommand{\fnum@figure}{\textbf{Figure~\thefigure}}

\usepackage{mathtools}
\newcommand{\g}[0]{\gamma}
\newcommand{\al}[0]{\alpha}
\newcommand{\be}[0]{\beta}
\newcommand{\de}[0]{\delta}

\newcommand{\dd}[0]{\partial}

\newcommand{\bs}[1]{\textbf{#1}}
\newcommand{\ea}[1]{\begin{align}#1\end{align}}
\newcommand{\eq}[1]{\begin{equation}#1\end{equation}}

\usepackage{cancel}

\begin{document}
\title{Predicting leptonic CP violation via minimization of neutrino entanglement}
\author{Gon\c{c}alo M. Quinta}
\email{goncalo.quinta@tecnico.ulisboa.pt}
\affiliation{Instituto de Telecomunica\c{c}\~{o}es, Lisboa, Portugal}

\author{Alexandre Sousa}
\email{alex.sousa@uc.edu}
\affiliation{Department of Physics, University of Cincinnati, Cincinnati, OH 45221, USA}

\author{Yasser Omar}
\email{yasser.omar@pqi.pt}
\affiliation{Instituto Superior Técnico, Universidade de Lisboa, Portugal}
\affiliation{Centro de Física e Engenharia de Materiais Avançados, Physics of Information and Quantum Technologies Group, Portugal}
\affiliation{Instituto de Telecomunica\c{c}\~{o}es, Lisboa, Portugal}
\affiliation{Portuguese Quantum Institute, Portugal}

%%%%%%%%%%%%%%%%%%%%%%%%%%%%%%%%%%%%%%%%%%%%%%%%%%%%%%%%%%%%%%%%%%%%%%%%%%%
%%%%%%%%%%%%%%%%%%%%%%%%%%%%%    Abstract     %%%%%%%%%%%%%%%%%%%%%%%%%%%%%
%%%%%%%%%%%%%%%%%%%%%%%%%%%%%%%%%%%%%%%%%%%%%%%%%%%%%%%%%%%%%%%%%%%%%%%%%%%

\begin{abstract}

We show how a minimization principle of quantum entanglement between the oscillating flavors of a neutrino leads to a unique prediction for the CP-violation phase in the neutrino sector without assuming extra symmetries in the Standard Model. We find a theoretical prediction consistent with either no CP-violation or a very small presence of it.

\end{abstract}

\maketitle

%%%%%%%%%%%%%%%%%%%%%%%%%%%%%%%%%%%%%%%%%%%%%%%%%%%%%%%%%%%%%%%%%%%%%%%%%%%
%%%%%%%%%%%%%%%%%%%%%%%%%%%    Introduction     %%%%%%%%%%%%%%%%%%%%%%%%%%%
%%%%%%%%%%%%%%%%%%%%%%%%%%%%%%%%%%%%%%%%%%%%%%%%%%%%%%%%%%%%%%%%%%%%%%%%%%%

The advent of quantum information (QI) has brought us a deeper understanding of quantum entanglement, a fundamental aspect of nature that Erwin Schrödinger called "the characteristic trait of quantum mechanics, the one that enforces its entire departure from classical lines of thought" \cite{schrodinger_1935}.  Furthermore, entanglement can also be seen from an operational point of view, namely for novel paradigms and disruptive technologies for information processing, information transmission, and information acquisition \cite{nielsen_chuang_2010}, with a significant potential for societal impacts. Quantum information science and technologies can, in turn, help us investigate, simulate, and better understand fundamental physics, such as quantum many-body systems or high-energy physics (HEP) \cite{snowmass_QC_HEP,PhysRevD.105.076012,snowmass_QC_HEP_2, Faulkner:2022mlp}. But what is exactly the role of quantum entanglement and quantum information in high-energy physics? Can quantum correlations help us understand and predict the behaviour of fundamental particles?

The Standard Model (SM) of particle physics has been extremely successful at compacting almost all of the subatomic physics. Nevertheless, most of its parameters are purely obtained via experiment band have been notoriously difficult to predict theoretically. Uncovering the fundamental origin of these parameters is of the utmost importance for our understanding of the physical world and may provide valuable insights towards a unifying theory of all interactions. The usual approach to this problem begins by assuming new ingredients in the form of symmetries, which typically add new particle content to the SM. Recently, however, there have been attempts at using quantum information to derive unexplained yet well established phenomena in the SM. These works usually involve finding specific situations in the SM which can be encoded in quantum states of up to three qubits. The most obvious cases consist in finding scenarios in particle physics where only two or three degrees of freedom interact. Helicities of fermionic particles (two degrees of freedom) and neutrino flavor oscillations (three degrees of freedom) are the most immediate ones. On neutrino oscillations, existing works (e.g. \cite{Blasone_2009, PhysRevD.82.093003, BLASONE2013320, Blasone_2014}) focus on calculating basic QI quantities, like entanglement entropy, verifying the possible entanglement classes or simulating the oscillation dynamics in a quantum computer. Regarding the helicities of fermionic particles, the analysis of the role of quantum scattering in entangling the helicity degrees of freedom of the participating particles has been explored in \cite{SciPostPhys.3.5.036, PhysRevD.100.105018}. Other works have also attempted to connect the symmetries of the SM with the entanglement classes of two and three qubits by treating Isopin as a qubit \cite{MULDERS2018193}. Overall, the only works that can be considered to constrain unexplained features of the SM are \cite{SciPostPhys.3.5.036}, where a principle of entanglement maximization was used to fix the symmetries of Quantum Electrodynamics (QED), and \cite{PhysRevLett.122.102001}, where a principle of entanglement suppression was proposed to explain emergent symmetries in Quantum Chromodynamics. Despite the growing interest in QI applications to the SM, however, the problem of predicting SM parameters with QI has never been addressed.

In this work, we will focus on one non-trivial example where entanglement is naturally occurring in particle physics, namely in the three quantum degrees of freedom associated to the flavor oscillations of a neutrino. Due to the mismatch between the mass and flavor basis, a non-interacting neutrino is always in a superposition of the three available flavor basis states, comprising a quantum state which is in general entangled. This entanglement is a function of the (currently unexplained) four parameters of the so-called Pontecorvo–Maki–Nakagawa–Sakata (PMNS) matrix, which quantifies the rotation between the mass and flavor eigenbases. In this work, we show how the minimisation of concurrence \cite{Po68}, a scalar monotonic quantity which quantifies the entanglement of a bi-partite quantum state, results in a unique prediction for the CP-violation phase.

%%%%%%%%%%%%%%%%%%%%%%%%%%%%%%%%%%%%%%%%%%%%%%%%%%%%%%%%%%%%%%%%%%%%%%%%%%%
%%%%%%%%%%%%%%%%%%%%    Entanglement in neutrinos     %%%%%%%%%%%%%%%%%%%%%
%%%%%%%%%%%%%%%%%%%%%%%%%%%%%%%%%%%%%%%%%%%%%%%%%%%%%%%%%%%%%%%%%%%%%%%%%%%

\textit{Entanglement in neutrino oscillations}--- Free neutrinos exist as mass eigenstates $\ket{\nu_i}$, where $i=1,2,3$ but they interact with matter in the flavor eigenbasis $\ket{\nu_\al}$, where $\al = e, \mu,\tau$. Henceforth, we will use latin indices for the mass eigenbasis and greek indices for the flavor eigenbasis. As a consequence, a neutrino with a given initial flavor $\al$ can also be written as the superposition
\eq{
\ket{\nu_\al} = \sum_{\al} U^{* \textrm{PMNS}}_{i \al} \ket{\nu_{i}}
}
where $U^{\textrm{PMNS}}$ is the PMNS matrix. The PMNS matrix has four degrees of freedom, and can be parametrized as
\begin{widetext}
\eq{
U^{\textrm{PMNS}} =
\begin{pmatrix}
1 & 0 & 0 \\
0 & \cos(\theta_{23}) & \sin(\theta_{23}) \\
0 & -\sin(\theta_{23}) & \cos(\theta_{23})
\end{pmatrix}
\begin{pmatrix}
\cos(\theta_{13}) & 0 & \sin(\theta_{13})e^{i \delta_{CP}} \\
0 & 1 & 0 \\
-\sin(\theta_{13})e^{-i \delta_{CP}} & \cos(\theta_{13}) & \cos(\theta_{13})
\end{pmatrix}
\begin{pmatrix}
\cos(\theta_{12}) & \sin(\theta_{12}) & 0 \\
-\sin(\theta_{12}) & \cos(\theta_{12}) & 0 \\
0 & 0 & 1
\end{pmatrix}\,,
}
\end{widetext}
which depends on three angles $\theta_{12},\theta_{13},\theta_{23}$ and a phase $\de_{CP}$. The latter is also known as the CP-violation phase, since it determines the violation of Charge-Parity (CP) invariance as quantified by a non-zero Jarslkog invariant, given by $J=\cos^2(\theta_{13})\sin(2\theta_{12})\sin(\theta_{13})\sin(2\theta_{23})\sin(\delta_{CP})/4$. After a time $t$, the initial state becomes $\ket{\nu_\al(t)} = \sum_{\al\be}U_{\al\be}(t)\ket{\nu_\be}$, where
\eq{\label{Utotal}
U_{\al\be}(t) = \sum_{i,j} U^{* \textrm{PMNS}}_{\al j} e^{-i E_j t} U^{\textrm{PMNS}}_{\be i}\,,
}
with $E_j = \sqrt{m_j^2+|\bs{p}|^2}$, encompasses the entire unitary evolution, given by a unitary rotation from the flavor to the mass eigenbasis and a free unitary evolution in the mass eigenbasis, followed by the inverse rotation back to the flavor eigenbasis. Due to the small masses of the neutrinos, one may take the approximation $E_j \approx |\bs{p}| + \frac{m^2_j}{2E}$ such that $e^{-i E_j t} \propto \exp(\textrm{diag}(0,-i \Delta m_{12}^2 \ell,-i \Delta m_{13}^2 \ell))$ up to an unphysical phase, where ${\Delta m_{ij}^2 = m^2_i-m^2_j}$ and $\ell \equiv t/(2E)$ is a parameter with units of distance. Since we are using natural units $c=\hbar=1$, one may use the time $t$ or distance $L$ interchangeably.

As a consequence of (\ref{Utotal}), a neutrino that started in a given flavor at time $t=0$, can be found to interact in a different flavor state after a time $t$, a phenomenon known as neutrino flavor oscillations. From the quantum information point of view, the superposition $\ket{\nu_\al(t)} = U_{\al e}(t) \ket{\nu_e} + U_{\al \mu}(t) \ket{\nu_{\mu}} + U_{\al \tau}(t) \ket{\nu_{\tau}}$ will in general result in an entangled state. To better emphasize the entangled degrees of freedom, one may take the mapping
\eq{\label{stoq}
\ket{\nu_e} = \ket{100}, \ket{\nu_{\mu}} = \ket{010}, \ket{\nu_{\tau}} = \ket{001},
}
which encodes flavors in three-qubit states. This is in fact the map used in most references dealing with entanglement in neutrino oscillations (see for example \cite{Blasone_2009,PhysRevD.82.093003,BLASONE2013320,Blasone_2014}). This mapping naturally emerges from a Quantum Field Theory perspective of neutrino oscillations \cite{Blasone_2014}, whereby each qubit represents the vacuum mode excitations corresponding to the creation and annihilation of quantum field modes associated to a flavor, i.e. $\ket{000} \equiv \ket{0}_{e}\otimes\ket{0}_{\mu}\otimes\ket{0}_{\tau}$, where $\ket{0}_{\al}$ is the vacuum for the flavor $\al$. Due to the fermionic nature of the neutrino quantum fields associated to each flavor, for a given momentum and spin there can only exist one excited mode per flavor, hence the qubit nature of the mapping. The fact that all the states in (\ref{stoq}) have only a single qubit in the state $\ket{1}$ is a consequence of the fact that only a single flavor per neutrino can be excited at the same time. This can be checked analytically by deriving the explicit form of the creation and annihilation operators associated to the mass and flavor neutrino quantum fields, as done in \cite{Blasone_2014}. In the end, the state, after a time $t$, of a neutrino with initial flavor $\al$ can be written as
\eq{\label{nuqubits}
\ket{\nu_\al(t)} = U_{\al e}(t) \ket{100} + U_{\al \mu}(t) \ket{010} + U_{\al \tau}(t) \ket{001}\,.
}
To study the entanglement in (\ref{nuqubits}), one may perform a Positive-Partial-Transpose (PTT) test \cite{HORODECKI19961} on the density matrix $\rho_{\al}(t) = \ket{\nu_\al(t)} \bra{\nu_\al(t)}$. This amounts to partial transposing one of the 3 qubits and calculating the eigenvalues of the resulting matrix. If one of them is negative, the matrix is entangled. One can check that indeed this is the case, as was already found in the literature \cite{Blasone_2009,PhysRevD.82.093003,BLASONE2013320,Blasone_2014}, and that the state $\ket{\nu_\al(t)}$ belongs to the W entanglement class of three qubits. In fact, it has already been verified that the entanglement present in the flavors degrees of freedom satisfies all monogamy inequalities \cite{Blasone_2009,PhysRevD.82.093003,BLASONE2013320,Blasone_2014}.

%%%%%%%%%%%%%%%%%%%%%%%%%%%%%%%%%%%%%%%%%%%%%%%%%%%%%%%%%%%%%%%%%%%%%%%%%%%
%%%%%%%%%%%%%%%%%%%%    Entanglement in neutrinos     %%%%%%%%%%%%%%%%%%%%%
%%%%%%%%%%%%%%%%%%%%%%%%%%%%%%%%%%%%%%%%%%%%%%%%%%%%%%%%%%%%%%%%%%%%%%%%%%%

\textit{A principle of entanglement minimisation}--- The residual tangle of (\ref{nuqubits}), a quantity which measures the amount of tripartite entanglement in a three qubit state \cite{tripartite}, is identically zero, as expected from a state of the W class. This implies that all entanglement exists among the three possible flavor bipartitions of $\ket{\nu_\al(t)}$. Bi-partite entanglement can be quantified by the squared concurrence \cite{PhysRevLett.78.5022} of a two-qubit state, in this case obtained by tracing out one qubit (i.e., flavor) from the density matrix $\rho_{\al}(t)$. One may show \cite{tripartite} that the squared concurrences for all bipartitions are given by
\ea{
C_{e\mu; \al}^2(\ell)    & = |U_{\al e}(\ell)|^2 |U_{\al \mu}(\ell)|^2 \,, \label{Cemu} \\
C_{e\tau; \al}^2(\ell)   & = |U_{\al e}(\ell)|^2 |U_{\al \tau}(\ell)|^2 \,, \label{Cetau} \\
C_{\mu\tau; \al}^2(\ell) & = |U_{\al \mu}(\ell)|^2 |U_{\al \tau}(\ell)|^2 \label{Cmutau} \,,
}
where $C_{\g\be; \al}^2$ denotes the squared concurrence between the two flavors $\g$ and $\be$, starting with an initial flavor $\al$ at $t=\ell=0$, where $\ell$ denotes the distance parameter $\ell \equiv t/(2E)$. The squared concurrences have involved expressions which depend on summations and products of trigonometric functions, resulting in oscillating quantities which are mostly intractable for analytic calculations. Nevertheless, these oscillating functions will have a certain amplitude, which depends on the four parameters $\theta_{12},\theta_{13},\theta_{23}$ and $\de_{CP}$ of the PMNS matrix, as well as the mass-squared differences $\Delta m^2_{21}$ and $\Delta m^2_{31}$. One may conjecture a principle of minimization of entanglement as a way to constrain the possible values of the free parameters of the PMNS matrix, similarly to the principle of least action. In this case, we assume the following:

\paragraph{{\bf Conjecture:}} The CP-violation phase $\delta_{CP}$ is such that the minimum range of bi-partite flavor entanglement is achieved.

%\begin{conjecture}\label{conjecture}
%The CP-violation phase $\delta_{CP}$ is such that the minimum range of bi-partite flavor %entanglement is achieved.
%\end{conjecture}

The reason why we consider only the $\delta_{CP}$ parameter is two-fold. Firstly, it is much simpler to minimize one parameter and to gain intuition on the problem. Secondly, the parameters $\theta_{12}, \theta_{13}$ and $\theta_{23}$ have been measured with a much better precision \cite{PDB} than $\delta_{CP}$, so the precision in the theoretical determination of $\delta_{CP}$ will be better as well. %In particular, throughout this work, the experimental values that will be used (written in the Particle Data Booklet form \cite{PhysRevD.101.116013}) are $\sin^2(\theta_{12})=0.307 \pm 0.013$,  $\sin^2(\theta_{13})=(2.20 \pm 0.07)\times 10^-2$, $\sin^2(\theta_{23})=0.539 \pm 0.022$ (for Inverted order) and $\sin^2(\theta_{23})=0.539 \pm 0.022$ (for Inverted order).

One may thus fix all experimental parameters except for $\delta_{CP}$ and calculate the values for which the minimum possible range among all squared concurrences is achieved. We begin by applying this minimization procedure on the pair of flavors $e$ and $\tau$, starting with an initial electron neutrino, so the squared concurrence $C_{e\tau; e}^2(\ell,\de_{CP})$ will be a function of $\ell$ and the phase $\de_{CP}$. Since $\delta_{CP}$ is not directly observable, we take the quantity $\sin(\delta_{CP})$ instead, which is observable via the Jarlskog invariant. The value of $\sin(\delta_{CP})$ for which the global maximum of $C_{e\tau; e}^2$ is minimized is obtained by solving the set of equations
\ea{
\frac{\dd C_{e\tau; e}^2(\ell,\de_{CP})}{\dd \ell}\bigg|_{\ell = \ell^{\textrm{max}}} & = 0 \,, \label{SE1} \\
\frac{\dd C_{e\tau; e}^2(\ell^{\textrm{max}},\de_{CP})}{\dd \sin(\delta_{CP})}\bigg|_{\sin(\delta_{CP}) = \sin(\delta_{CP})^{\textrm{min}}} & = 0\,. \label{SE2}
}
Although the system of equations given by (\ref{SE1}) and (\ref{SE2}) is too intricate to solve analytically, one may plot the results in order to get a good estimate. As one may immediately infer from Figure 1, the numerical solution for $\sin(\delta_{CP})^{\textrm{min}}$ is quite close to $0$,
\begin{figure}[h!]
\includegraphics[width=\linewidth]{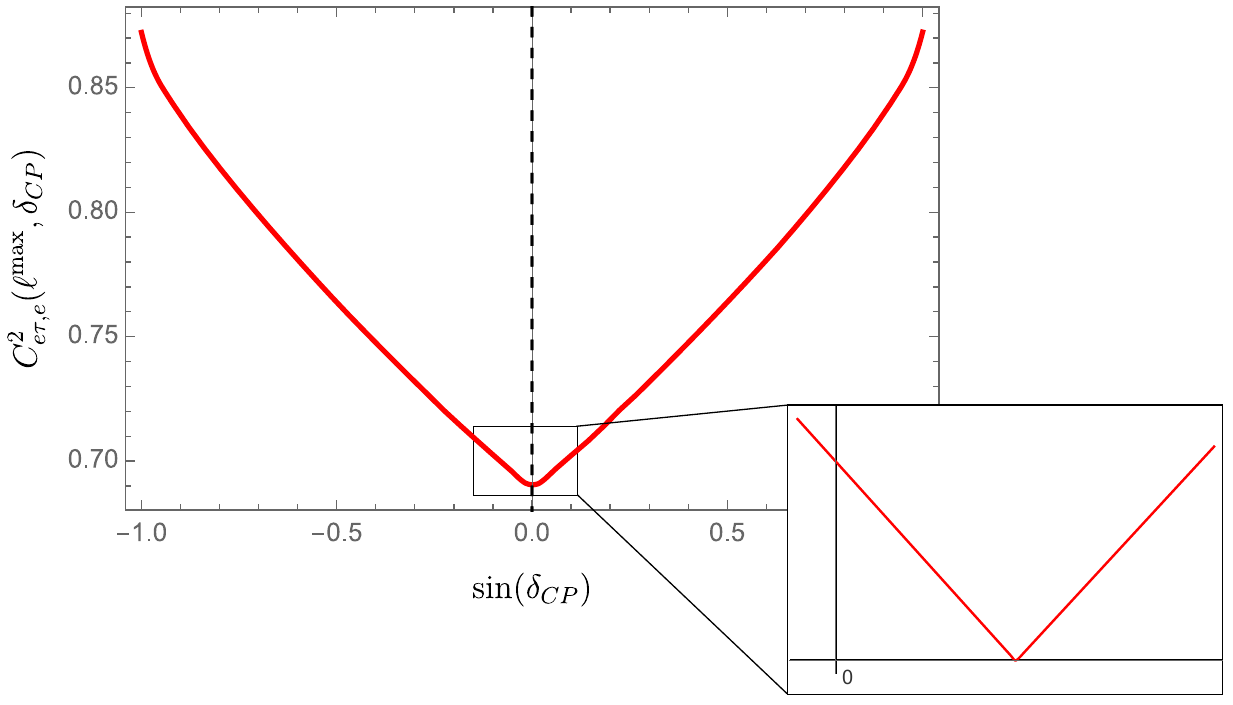}
\caption{Numerical solution of equations (\ref{SE1}) and (\ref{SE2}) with respect to $\sin(\delta_{CP})$. The global minimum is unique and approximately equal to $\sin(\delta_{CP}) \approx 0.000474$. All free parameters apart from $\sin(\delta_{CP})$ are fixed according to the most recent experimental data from the Particle Data Book~\cite{PDB}, using 1-sigma errors.}
\label{minimum}
\end{figure}
which is not so intuitive, given the involved form of $C_{e\tau; \al}^2$, as can be observed in Figure 2. In addition, it is also surprising that there is a single value of $\sin(\delta_{CP})$ for which the system of equations (\ref{SE1}) and (\ref{SE2}) is satisfied, which is not expected given the oscillating nature of $C_{e\tau; e}^2$. Nevertheless, with this ansatz, one may check to a high precision that the actual minimum is achieved for $\sin(\delta_{CP})^{\textrm{min}} \approx 0.000474$.
\begin{figure}[h!]
\includegraphics[width=\linewidth]{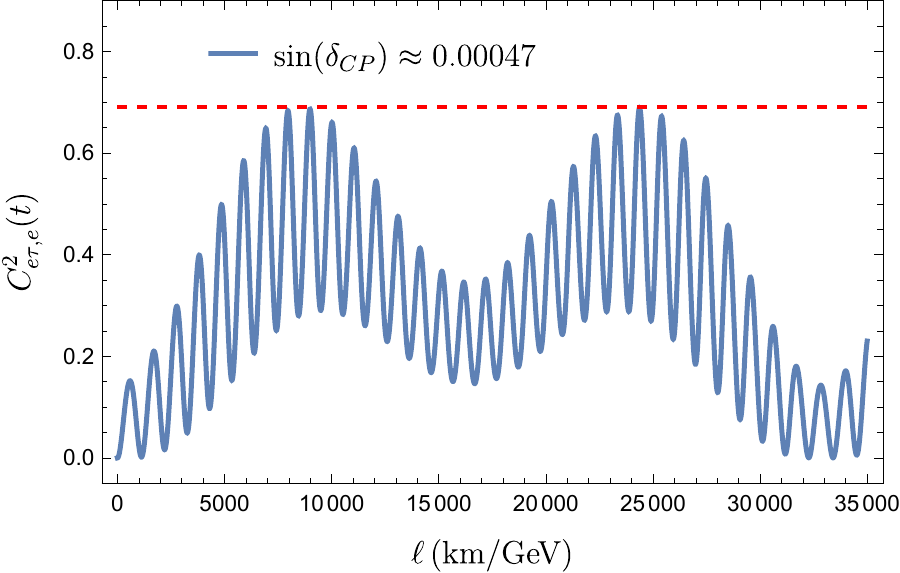}
\caption{Concurrence $C_{e\tau; e}^2(\ell,\de_{CP})$ (in blue) as a function of the parameter $\ell$, for the specific value of $\sin(\delta_{CP}) \approx 0.000474$. This corresponds to the scenario where the minimum range of entanglement between the electron and muon flavors is achieved, i.e. the minimum of the possible maxima (in red) as a function of $\sin(\delta_{CP})$ is achieved. All free parameters apart from $\sin(\delta_{CP})$ are fixed according to the most recent experimental data \cite{PDB}.}
\label{oscillations}
\end{figure}

Repeating the same procedure for all other possible pairs of flavors and initial neutrino flavors, one finds the results plotted in Figure 3, which confirm that the minimum possible range of entanglement happens for the bipartition of electron/tau flavors.
\begin{figure}[h!]
\includegraphics[width=\linewidth]{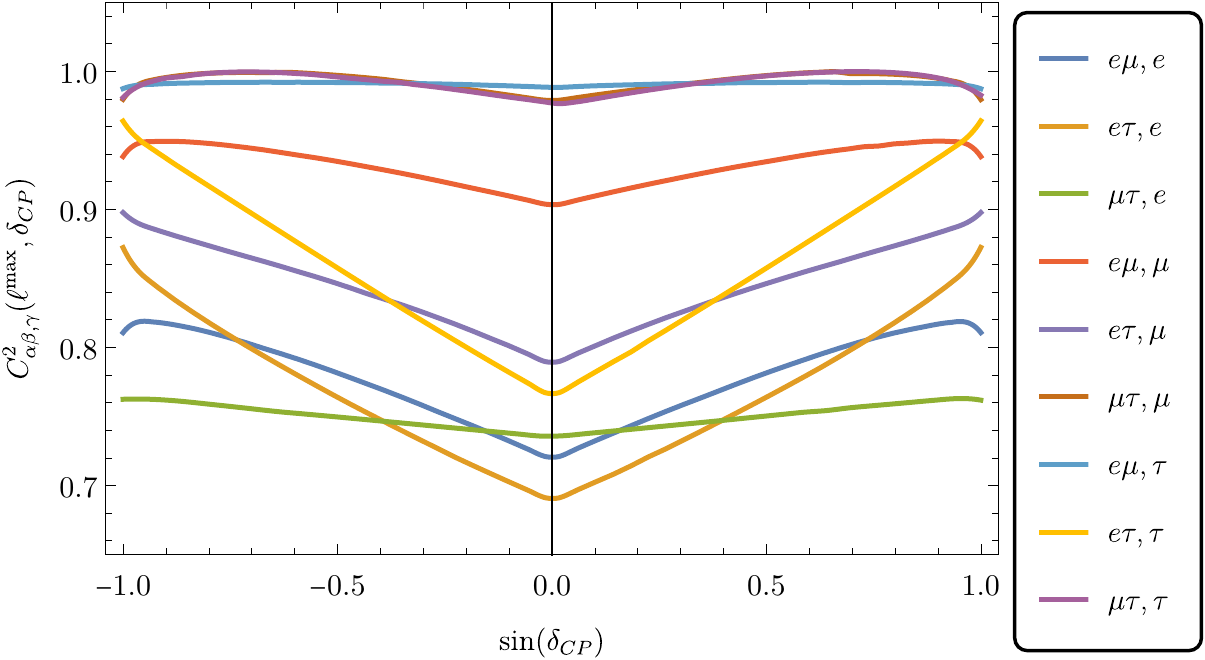}
\caption{The concurrences between all possible pairs of neutrino flavors, for all initial neutrino flavors possibilities. It is found numerically that the global minimum of all these functions is achieved for the electron/tau neutrino flavor case, starting with an initial electron neutrino. All free parameters apart from $\delta_{CP}$ are fixed according to the most recent experimental data \cite{PDB}.}
\label{oscillations2}
\end{figure}
Due to the non-linear nature of the concurrence oscillations, it is difficult to find any intuition in the magnitudes for the maximum ranges present in Fig.~\ref{oscillations2}. Nevertheless, that fact that $\theta_{13}$ is the smallest mixing angle could be the reason why the concurrence for the bipartition $e/\tau$ with the electron flavored neutrino has the smallest minimum range. Checking individually the range of each concurrence for all pairs of flavors, one would find that for some cases the minimum would be close to 0 as well, although the correction to this value compared to the electron/tau case would be different. Nevertheless, since the latter is the minimum possible range of entanglement, one is led to the prediction
\eq{\label{NOcp}
\sin(\de^{\textrm{NO}}_{CP}) = 4.7^{+81.1}_{-27.5} \times 10^{-4} \,,
}
for the Normal Ordering of neutrino masses and
\eq{\label{IOcp}
\sin(\de^{\textrm{IO}}_{CP}) = -2.1^{+39.5}_{-84.6} \times 10^{-4} \,,
}
for the Inverted Ordering. These results are obtained using the central values and experimental errors in the Particle Data Book~\cite{PDB} and are contained within three standard deviations from the current experimental average $\de^{\textrm{exp.}}_{CP} - \pi = 0.36^{+0.20}_{0.16} \, \pi$, or equivalently, $\sin(\de^{\textrm{exp.}}_{CP}) = -0.904^{+0.077}_{-0.317}$. Our results, expressed in Eqs.~(\ref{NOcp})-(\ref{IOcp}), thus favor CP-conservation in the lepton sector.

%%%%%%%%%%%%%%%%%%%%%%%%%%%%%%%%%%%%%%%%%%%%%%%%%%%%%%%%%%%%%%%%%%%%%%%%%%%
%%%%%%%%%%%%%%%%%%%%%%%%%%%%    Conclusions     %%%%%%%%%%%%%%%%%%%%%%%%%%%
%%%%%%%%%%%%%%%%%%%%%%%%%%%%%%%%%%%%%%%%%%%%%%%%%%%%%%%%%%%%%%%%%%%%%%%%%%%

\textit{Conclusions}--- In this work we showed how one may formulate principles that minimize entanglement, and how this may be used as a condition to determine unexplained parameters in the Standard Model of Particle Physics. In particular, the maximum concurrence between two flavors of a neutrino, which quantifies the entanglement between those two degrees of freedom, was minimized with respect to the CP-violation phase. This resulted in a unique prediction to the latter parameter which, despite being too difficult to achieve analytically, resulted nonetheless in concrete values with a small margin of error, favoring CP-conservation. This result opens the prospect of exploiting quantum information, and in particular quantum correlations to understand and predict the behaviour of fundamental particles.

%%%%%%%%%%%%%%%%%%%%%%%%%%%%%%%%%%%%%%%%%%%%%%%%%%%%%%%%%%%%%%%%%%%%%%%%%%%
%%%%%%%%%%%%%%%%%%%%%%%%%    Acknowledgements     %%%%%%%%%%%%%%%%%%%%%%%%%
%%%%%%%%%%%%%%%%%%%%%%%%%%%%%%%%%%%%%%%%%%%%%%%%%%%%%%%%%%%%%%%%%%%%%%%%%%%

\textit{Acknowledgements}--- The authors thank Duarte Magano, João Penedo, João Seixas and Pedram Bargassa for fruitful discussions. GQ and YO thank the support from project QuantHEP – Quantum Computing Solutions for High-Energy Physics, supported by the EU H2020 QuantERA ERA-NET Cofund in Quantum Technologies and FCT -- Funda\c{c}\~{a}o para a Ci\^{e}ncia e a Tecnologia (QuantERA/0001/2019), as well as the support from FCT through projects UIDB/50008/2020, UIDB/04540/2020, CEECIND/02474/2018 and EXPL/FIS-PAR/1604/2021 QEntHEP - Quantum Entanglement in High Energy Physics.

%%%%%%%%%%%%%%%%%%%%%%%%%%%%%%%%%%%%%%%%%%%%%%%%%%%%%%%%%%%%%%%%%%%%%%%%%%%
%%%%%%%%%%%%%%%%%%%%%%%%%%%    Bibliography     %%%%%%%%%%%%%%%%%%%%%%%%%%%
%%%%%%%%%%%%%%%%%%%%%%%%%%%%%%%%%%%%%%%%%%%%%%%%%%%%%%%%%%%%%%%%%%%%%%%%%%%

\bibliography{Physics.bib}

\begin{thebibliography}{19}%
\makeatletter
\providecommand \@ifxundefined [1]{%
 \@ifx{#1\undefined}
}%
\providecommand \@ifnum [1]{%
 \ifnum #1\expandafter \@firstoftwo
 \else \expandafter \@secondoftwo
 \fi
}%
\providecommand \@ifx [1]{%
 \ifx #1\expandafter \@firstoftwo
 \else \expandafter \@secondoftwo
 \fi
}%
\providecommand \natexlab [1]{#1}%
\providecommand \enquote  [1]{``#1''}%
\providecommand \bibnamefont  [1]{#1}%
\providecommand \bibfnamefont [1]{#1}%
\providecommand \citenamefont [1]{#1}%
\providecommand \href@noop [0]{\@secondoftwo}%
\providecommand \href [0]{\begingroup \@sanitize@url \@href}%
\providecommand \@href[1]{\@@startlink{#1}\@@href}%
\providecommand \@@href[1]{\endgroup#1\@@endlink}%
\providecommand \@sanitize@url [0]{\catcode `\\12\catcode `\$12\catcode
  `\&12\catcode `\#12\catcode `\^12\catcode `\_12\catcode `\%12\relax}%
\providecommand \@@startlink[1]{}%
\providecommand \@@endlink[0]{}%
\providecommand \url  [0]{\begingroup\@sanitize@url \@url }%
\providecommand \@url [1]{\endgroup\@href {#1}{\urlprefix }}%
\providecommand \urlprefix  [0]{URL }%
\providecommand \Eprint [0]{\href }%
\providecommand \doibase [0]{https://doi.org/}%
\providecommand \selectlanguage [0]{\@gobble}%
\providecommand \bibinfo  [0]{\@secondoftwo}%
\providecommand \bibfield  [0]{\@secondoftwo}%
\providecommand \translation [1]{[#1]}%
\providecommand \BibitemOpen [0]{}%
\providecommand \bibitemStop [0]{}%
\providecommand \bibitemNoStop [0]{.\EOS\space}%
\providecommand \EOS [0]{\spacefactor3000\relax}%
\providecommand \BibitemShut  [1]{\csname bibitem#1\endcsname}%
\let\auto@bib@innerbib\@empty
%</preamble>
\bibitem [{\citenamefont {Schrödinger}(1935)}]{schrodinger_1935}%
  \BibitemOpen
  \bibfield  {author} {\bibinfo {author} {\bibfnamefont {E.}~\bibnamefont
  {Schrödinger}},\ }\bibfield  {title} {\bibinfo {title} {Discussion of
  probability relations between separated systems},\ }\href
  {https://doi.org/10.1017/S0305004100013554} {\bibfield  {journal} {\bibinfo
  {journal} {Mathematical Proceedings of the Cambridge Philosophical Society}\
  }\textbf {\bibinfo {volume} {31}},\ \bibinfo {pages} {555–563} (\bibinfo
  {year} {1935})}\BibitemShut {NoStop}%
\bibitem [{\citenamefont {Nielsen}\ and\ \citenamefont
  {Chuang}(2010)}]{nielsen_chuang_2010}%
  \BibitemOpen
  \bibfield  {author} {\bibinfo {author} {\bibfnamefont {M.~A.}\ \bibnamefont
  {Nielsen}}\ and\ \bibinfo {author} {\bibfnamefont {I.~L.}\ \bibnamefont
  {Chuang}},\ }\href {https://doi.org/10.1017/CBO9780511976667} {\emph
  {\bibinfo {title} {Quantum Computation and Quantum Information: 10th
  Anniversary Edition}}}\ (\bibinfo  {publisher} {Cambridge University Press},\
  \bibinfo {year} {2010})\BibitemShut {NoStop}%
\bibitem [{\citenamefont {Humble}\ \emph {et~al.}(2022)\citenamefont {Humble},
  \citenamefont {Delgado}, \citenamefont {Pooser}, \citenamefont {Seck},
  \citenamefont {Bennink}, \citenamefont {Leyton-Ortega}, \citenamefont {Wang},
  \citenamefont {Dumitrescu}, \citenamefont {Morris}, \citenamefont {Hamilton},
  \citenamefont {Lyakh}, \citenamefont {Date}, \citenamefont {Wang},
  \citenamefont {Peters}, \citenamefont {Evans}, \citenamefont {Demarteau},
  \citenamefont {McCaskey}, \citenamefont {Nguyen}, \citenamefont {Clark},
  \citenamefont {Reville}, \citenamefont {Di~Meglio}, \citenamefont {Grossi},
  \citenamefont {Vallecorsa}, \citenamefont {Borras}, \citenamefont {Jansen},\
  and\ \citenamefont {Krücker}}]{snowmass_QC_HEP}%
  \BibitemOpen
  \bibfield  {author} {\bibinfo {author} {\bibfnamefont {T.~S.}\ \bibnamefont
  {Humble}}, \bibinfo {author} {\bibfnamefont {A.}~\bibnamefont {Delgado}},
  \bibinfo {author} {\bibfnamefont {R.}~\bibnamefont {Pooser}}, \bibinfo
  {author} {\bibfnamefont {C.}~\bibnamefont {Seck}}, \bibinfo {author}
  {\bibfnamefont {R.}~\bibnamefont {Bennink}}, \bibinfo {author} {\bibfnamefont
  {V.}~\bibnamefont {Leyton-Ortega}}, \bibinfo {author} {\bibfnamefont
  {C.~C.~J.}\ \bibnamefont {Wang}}, \bibinfo {author} {\bibfnamefont
  {E.}~\bibnamefont {Dumitrescu}}, \bibinfo {author} {\bibfnamefont
  {T.}~\bibnamefont {Morris}}, \bibinfo {author} {\bibfnamefont
  {K.}~\bibnamefont {Hamilton}}, \bibinfo {author} {\bibfnamefont
  {D.}~\bibnamefont {Lyakh}}, \bibinfo {author} {\bibfnamefont
  {P.}~\bibnamefont {Date}}, \bibinfo {author} {\bibfnamefont {Y.}~\bibnamefont
  {Wang}}, \bibinfo {author} {\bibfnamefont {N.~A.}\ \bibnamefont {Peters}},
  \bibinfo {author} {\bibfnamefont {K.~J.}\ \bibnamefont {Evans}}, \bibinfo
  {author} {\bibfnamefont {M.}~\bibnamefont {Demarteau}}, \bibinfo {author}
  {\bibfnamefont {A.}~\bibnamefont {McCaskey}}, \bibinfo {author}
  {\bibfnamefont {T.}~\bibnamefont {Nguyen}}, \bibinfo {author} {\bibfnamefont
  {S.}~\bibnamefont {Clark}}, \bibinfo {author} {\bibfnamefont
  {M.}~\bibnamefont {Reville}}, \bibinfo {author} {\bibfnamefont
  {A.}~\bibnamefont {Di~Meglio}}, \bibinfo {author} {\bibfnamefont
  {M.}~\bibnamefont {Grossi}}, \bibinfo {author} {\bibfnamefont
  {S.}~\bibnamefont {Vallecorsa}}, \bibinfo {author} {\bibfnamefont
  {K.}~\bibnamefont {Borras}}, \bibinfo {author} {\bibfnamefont
  {K.}~\bibnamefont {Jansen}},\ and\ \bibinfo {author} {\bibfnamefont
  {D.}~\bibnamefont {Krücker}},\ }\href
  {https://doi.org/10.48550/ARXIV.2203.07091} {\bibinfo {title} {Snowmass white
  paper: Quantum computing systems and software for high-energy physics
  research}} (\bibinfo {year} {2022})\BibitemShut {NoStop}%
\bibitem [{\citenamefont {Magano}\ \emph {et~al.}(2022)\citenamefont {Magano},
  \citenamefont {Kumar}, \citenamefont {K\ifmmode~\bar{a}\else \={a}\fi{}lis},
  \citenamefont {Loc\ifmmode~\bar{a}\else \={a}\fi{}ns}, \citenamefont {Glos},
  \citenamefont {Pratapsi}, \citenamefont {Quinta}, \citenamefont
  {Dimitrijevs}, \citenamefont {Rivo\ifmmode~\check{s}\else \v{s}\fi{}s},
  \citenamefont {Bargassa}, \citenamefont {Seixas}, \citenamefont {Ambainis},\
  and\ \citenamefont {Omar}}]{PhysRevD.105.076012}%
  \BibitemOpen
  \bibfield  {author} {\bibinfo {author} {\bibfnamefont {D.}~\bibnamefont
  {Magano}}, \bibinfo {author} {\bibfnamefont {A.}~\bibnamefont {Kumar}},
  \bibinfo {author} {\bibfnamefont {M.}~\bibnamefont {K\ifmmode~\bar{a}\else
  \={a}\fi{}lis}}, \bibinfo {author} {\bibfnamefont {A.}~\bibnamefont
  {Loc\ifmmode~\bar{a}\else \={a}\fi{}ns}}, \bibinfo {author} {\bibfnamefont
  {A.}~\bibnamefont {Glos}}, \bibinfo {author} {\bibfnamefont {S.}~\bibnamefont
  {Pratapsi}}, \bibinfo {author} {\bibfnamefont {G.}~\bibnamefont {Quinta}},
  \bibinfo {author} {\bibfnamefont {M.}~\bibnamefont {Dimitrijevs}}, \bibinfo
  {author} {\bibfnamefont {A.}~\bibnamefont {Rivo\ifmmode~\check{s}\else
  \v{s}\fi{}s}}, \bibinfo {author} {\bibfnamefont {P.}~\bibnamefont
  {Bargassa}}, \bibinfo {author} {\bibfnamefont {J.}~\bibnamefont {Seixas}},
  \bibinfo {author} {\bibfnamefont {A.}~\bibnamefont {Ambainis}},\ and\
  \bibinfo {author} {\bibfnamefont {Y.}~\bibnamefont {Omar}},\ }\bibfield
  {title} {\bibinfo {title} {Quantum speedup for track reconstruction in
  particle accelerators},\ }\href {https://doi.org/10.1103/PhysRevD.105.076012}
  {\bibfield  {journal} {\bibinfo  {journal} {Phys. Rev. D}\ }\textbf {\bibinfo
  {volume} {105}},\ \bibinfo {pages} {076012} (\bibinfo {year}
  {2022})}\BibitemShut {NoStop}%
\bibitem [{\citenamefont {Delgado}\ \emph {et~al.}(2022)\citenamefont
  {Delgado}, \citenamefont {Hamilton}, \citenamefont {Date}, \citenamefont
  {Vlimant}, \citenamefont {Magano}, \citenamefont {Omar}, \citenamefont
  {Bargassa}, \citenamefont {Francis}, \citenamefont {Gianelle}, \citenamefont
  {Sestini}, \citenamefont {Lucchesi}, \citenamefont {Zuliani}, \citenamefont
  {Nicotra}, \citenamefont {de~Vries}, \citenamefont {Dibenedetto},
  \citenamefont {Martinez}, \citenamefont {Rodrigues}, \citenamefont {Sierra},
  \citenamefont {Vallecorsa}, \citenamefont {Thaler}, \citenamefont
  {Bravo-Prieto}, \citenamefont {Chang}, \citenamefont {Lazar},\ and\
  \citenamefont {Argüelles}}]{snowmass_QC_HEP_2}%
  \BibitemOpen
  \bibfield  {author} {\bibinfo {author} {\bibfnamefont {A.}~\bibnamefont
  {Delgado}}, \bibinfo {author} {\bibfnamefont {K.~E.}\ \bibnamefont
  {Hamilton}}, \bibinfo {author} {\bibfnamefont {P.}~\bibnamefont {Date}},
  \bibinfo {author} {\bibfnamefont {J.-R.}\ \bibnamefont {Vlimant}}, \bibinfo
  {author} {\bibfnamefont {D.}~\bibnamefont {Magano}}, \bibinfo {author}
  {\bibfnamefont {Y.}~\bibnamefont {Omar}}, \bibinfo {author} {\bibfnamefont
  {P.}~\bibnamefont {Bargassa}}, \bibinfo {author} {\bibfnamefont
  {A.}~\bibnamefont {Francis}}, \bibinfo {author} {\bibfnamefont
  {A.}~\bibnamefont {Gianelle}}, \bibinfo {author} {\bibfnamefont
  {L.}~\bibnamefont {Sestini}}, \bibinfo {author} {\bibfnamefont
  {D.}~\bibnamefont {Lucchesi}}, \bibinfo {author} {\bibfnamefont
  {D.}~\bibnamefont {Zuliani}}, \bibinfo {author} {\bibfnamefont
  {D.}~\bibnamefont {Nicotra}}, \bibinfo {author} {\bibfnamefont
  {J.}~\bibnamefont {de~Vries}}, \bibinfo {author} {\bibfnamefont
  {D.}~\bibnamefont {Dibenedetto}}, \bibinfo {author} {\bibfnamefont {M.~L.}\
  \bibnamefont {Martinez}}, \bibinfo {author} {\bibfnamefont {E.}~\bibnamefont
  {Rodrigues}}, \bibinfo {author} {\bibfnamefont {C.~V.}\ \bibnamefont
  {Sierra}}, \bibinfo {author} {\bibfnamefont {S.}~\bibnamefont {Vallecorsa}},
  \bibinfo {author} {\bibfnamefont {J.}~\bibnamefont {Thaler}}, \bibinfo
  {author} {\bibfnamefont {C.}~\bibnamefont {Bravo-Prieto}}, \bibinfo {author}
  {\bibfnamefont {s.~Y.}\ \bibnamefont {Chang}}, \bibinfo {author}
  {\bibfnamefont {J.}~\bibnamefont {Lazar}},\ and\ \bibinfo {author}
  {\bibfnamefont {C.~A.}\ \bibnamefont {Argüelles}},\ }\href
  {https://doi.org/10.48550/ARXIV.2203.08805} {\bibinfo {title} {Quantum
  computing for data analysis in high-energy physics}} (\bibinfo {year}
  {2022})\BibitemShut {NoStop}%
\bibitem [{\citenamefont {Faulkner}\ \emph {et~al.}(2022)\citenamefont
  {Faulkner}, \citenamefont {Hartman}, \citenamefont {Headrick}, \citenamefont
  {Rangamani},\ and\ \citenamefont {Swingle}}]{Faulkner:2022mlp}%
  \BibitemOpen
  \bibfield  {author} {\bibinfo {author} {\bibfnamefont {T.}~\bibnamefont
  {Faulkner}}, \bibinfo {author} {\bibfnamefont {T.}~\bibnamefont {Hartman}},
  \bibinfo {author} {\bibfnamefont {M.}~\bibnamefont {Headrick}}, \bibinfo
  {author} {\bibfnamefont {M.}~\bibnamefont {Rangamani}},\ and\ \bibinfo
  {author} {\bibfnamefont {B.}~\bibnamefont {Swingle}},\ }\bibfield  {title}
  {\bibinfo {title} {{Snowmass white paper: Quantum information in quantum
  field theory and quantum gravity}},\ }in\ \href@noop {} {\emph {\bibinfo
  {booktitle} {{2022 Snowmass Summer Study}}}}\ (\bibinfo {year} {2022})\
  \Eprint {https://arxiv.org/abs/2203.07117} {arXiv:2203.07117 [hep-th]}
  \BibitemShut {NoStop}%
\bibitem [{\citenamefont {Blasone}\ \emph {et~al.}(2009)\citenamefont
  {Blasone}, \citenamefont {Dell{\textquotesingle}Anno}, \citenamefont
  {Siena},\ and\ \citenamefont {Illuminati}}]{Blasone_2009}%
  \BibitemOpen
  \bibfield  {author} {\bibinfo {author} {\bibfnamefont {M.}~\bibnamefont
  {Blasone}}, \bibinfo {author} {\bibfnamefont {F.}~\bibnamefont
  {Dell{\textquotesingle}Anno}}, \bibinfo {author} {\bibfnamefont {S.~D.}\
  \bibnamefont {Siena}},\ and\ \bibinfo {author} {\bibfnamefont
  {F.}~\bibnamefont {Illuminati}},\ }\bibfield  {title} {\bibinfo {title}
  {Entanglement in neutrino oscillations},\ }\href
  {https://doi.org/10.1209/0295-5075/85/50002} {\bibfield  {journal} {\bibinfo
  {journal} {{EPL} (Europhysics Letters)}\ }\textbf {\bibinfo {volume} {85}},\
  \bibinfo {pages} {50002} (\bibinfo {year} {2009})}\BibitemShut {NoStop}%
\bibitem [{\citenamefont {Kayser}\ \emph {et~al.}(2010)\citenamefont {Kayser},
  \citenamefont {Kopp}, \citenamefont {Robertson},\ and\ \citenamefont
  {Vogel}}]{PhysRevD.82.093003}%
  \BibitemOpen
  \bibfield  {author} {\bibinfo {author} {\bibfnamefont {B.}~\bibnamefont
  {Kayser}}, \bibinfo {author} {\bibfnamefont {J.}~\bibnamefont {Kopp}},
  \bibinfo {author} {\bibfnamefont {R.~G.~H.}\ \bibnamefont {Robertson}},\ and\
  \bibinfo {author} {\bibfnamefont {P.}~\bibnamefont {Vogel}},\ }\bibfield
  {title} {\bibinfo {title} {Theory of neutrino oscillations with
  entanglement},\ }\href {https://doi.org/10.1103/PhysRevD.82.093003}
  {\bibfield  {journal} {\bibinfo  {journal} {Phys. Rev. D}\ }\textbf {\bibinfo
  {volume} {82}},\ \bibinfo {pages} {093003} (\bibinfo {year}
  {2010})}\BibitemShut {NoStop}%
\bibitem [{BLA(2013)}]{BLASONE2013320}%
  \BibitemOpen
  \bibfield  {title} {\bibinfo {title} {Neutrino flavor entanglement},\ }\href
  {https://doi.org/https://doi.org/10.1016/j.nuclphysbps.2013.04.116}
  {\bibfield  {journal} {\bibinfo  {journal} {Nuclear Physics B - Proceedings
  Supplements}\ }\textbf {\bibinfo {volume} {237-238}},\ \bibinfo {pages} {320}
  (\bibinfo {year} {2013})},\ \bibinfo {note} {proceedings of the Neutrino
  Oscillation Workshop}\BibitemShut {NoStop}%
\bibitem [{\citenamefont {Blasone}\ \emph {et~al.}(2014)\citenamefont
  {Blasone}, \citenamefont {Dell{\textquotesingle}Anno}, \citenamefont
  {Siena},\ and\ \citenamefont {Illuminati}}]{Blasone_2014}%
  \BibitemOpen
  \bibfield  {author} {\bibinfo {author} {\bibfnamefont {M.}~\bibnamefont
  {Blasone}}, \bibinfo {author} {\bibfnamefont {F.}~\bibnamefont
  {Dell{\textquotesingle}Anno}}, \bibinfo {author} {\bibfnamefont {S.~D.}\
  \bibnamefont {Siena}},\ and\ \bibinfo {author} {\bibfnamefont
  {F.}~\bibnamefont {Illuminati}},\ }\bibfield  {title} {\bibinfo {title} {A
  field-theoretical approach to entanglement in neutrino mixing and
  oscillations},\ }\href {https://doi.org/10.1209/0295-5075/106/30002}
  {\bibfield  {journal} {\bibinfo  {journal} {{EPL} (Europhysics Letters)}\
  }\textbf {\bibinfo {volume} {106}},\ \bibinfo {pages} {30002} (\bibinfo
  {year} {2014})}\BibitemShut {NoStop}%
\bibitem [{\citenamefont {Cervera-Lierta}\ \emph {et~al.}(2017)\citenamefont
  {Cervera-Lierta}, \citenamefont {Latorre}, \citenamefont {Rojo},\ and\
  \citenamefont {Rottoli}}]{SciPostPhys.3.5.036}%
  \BibitemOpen
  \bibfield  {author} {\bibinfo {author} {\bibfnamefont {A.}~\bibnamefont
  {Cervera-Lierta}}, \bibinfo {author} {\bibfnamefont {J.~I.}\ \bibnamefont
  {Latorre}}, \bibinfo {author} {\bibfnamefont {J.}~\bibnamefont {Rojo}},\ and\
  \bibinfo {author} {\bibfnamefont {L.}~\bibnamefont {Rottoli}},\ }\bibfield
  {title} {\bibinfo {title} {{Maximal Entanglement in High Energy Physics}},\
  }\href {https://doi.org/10.21468/SciPostPhys.3.5.036} {\bibfield  {journal}
  {\bibinfo  {journal} {SciPost Phys.}\ }\textbf {\bibinfo {volume} {3}},\
  \bibinfo {pages} {036} (\bibinfo {year} {2017})}\BibitemShut {NoStop}%
\bibitem [{\citenamefont {Araujo}\ \emph {et~al.}(2019)\citenamefont {Araujo},
  \citenamefont {Hiller}, \citenamefont {da~Paz}, \citenamefont {Ferreira},
  \citenamefont {Sampaio},\ and\ \citenamefont {Costa}}]{PhysRevD.100.105018}%
  \BibitemOpen
  \bibfield  {author} {\bibinfo {author} {\bibfnamefont {J.~B.}\ \bibnamefont
  {Araujo}}, \bibinfo {author} {\bibfnamefont {B.}~\bibnamefont {Hiller}},
  \bibinfo {author} {\bibfnamefont {I.~G.}\ \bibnamefont {da~Paz}}, \bibinfo
  {author} {\bibfnamefont {M.~M.}\ \bibnamefont {Ferreira}}, \bibinfo {author}
  {\bibfnamefont {M.}~\bibnamefont {Sampaio}},\ and\ \bibinfo {author}
  {\bibfnamefont {H.~A.~S.}\ \bibnamefont {Costa}},\ }\bibfield  {title}
  {\bibinfo {title} {Measuring qed cross sections via entanglement},\ }\href
  {https://doi.org/10.1103/PhysRevD.100.105018} {\bibfield  {journal} {\bibinfo
   {journal} {Phys. Rev. D}\ }\textbf {\bibinfo {volume} {100}},\ \bibinfo
  {pages} {105018} (\bibinfo {year} {2019})}\BibitemShut {NoStop}%
\bibitem [{\citenamefont {Mulders}(2018)}]{MULDERS2018193}%
  \BibitemOpen
  \bibfield  {author} {\bibinfo {author} {\bibfnamefont {P.}~\bibnamefont
  {Mulders}},\ }\bibfield  {title} {\bibinfo {title} {Emergent symmetries of
  the standard model},\ }\href
  {https://doi.org/https://doi.org/10.1016/j.physletb.2018.09.063} {\bibfield
  {journal} {\bibinfo  {journal} {Physics Letters B}\ }\textbf {\bibinfo
  {volume} {787}},\ \bibinfo {pages} {193} (\bibinfo {year}
  {2018})}\BibitemShut {NoStop}%
\bibitem [{\citenamefont {Beane}\ \emph {et~al.}(2019)\citenamefont {Beane},
  \citenamefont {Kaplan}, \citenamefont {Klco},\ and\ \citenamefont
  {Savage}}]{PhysRevLett.122.102001}%
  \BibitemOpen
  \bibfield  {author} {\bibinfo {author} {\bibfnamefont {S.~R.}\ \bibnamefont
  {Beane}}, \bibinfo {author} {\bibfnamefont {D.~B.}\ \bibnamefont {Kaplan}},
  \bibinfo {author} {\bibfnamefont {N.}~\bibnamefont {Klco}},\ and\ \bibinfo
  {author} {\bibfnamefont {M.~J.}\ \bibnamefont {Savage}},\ }\bibfield  {title}
  {\bibinfo {title} {Entanglement suppression and emergent symmetries of strong
  interactions},\ }\href {https://doi.org/10.1103/PhysRevLett.122.102001}
  {\bibfield  {journal} {\bibinfo  {journal} {Phys. Rev. Lett.}\ }\textbf
  {\bibinfo {volume} {122}},\ \bibinfo {pages} {102001} (\bibinfo {year}
  {2019})}\BibitemShut {NoStop}%
\bibitem [{\citenamefont {Pontecorvo}(1968)}]{Po68}%
  \BibitemOpen
  \bibfield  {author} {\bibinfo {author} {\bibfnamefont {B.}~\bibnamefont
  {Pontecorvo}},\ }\bibfield  {title} {\bibinfo {title} {Neutrino experiments
  and the problem of conservation of leptonic charge},\ }\href@noop {}
  {\bibfield  {journal} {\bibinfo  {journal} {Soviet Physics JETP}\ }\textbf
  {\bibinfo {volume} {26}},\ \bibinfo {pages} {984} (\bibinfo {year}
  {1968})}\BibitemShut {NoStop}%
\bibitem [{HOR(1996)}]{HORODECKI19961}%
  \BibitemOpen
  \bibfield  {title} {\bibinfo {title} {Separability of mixed states: necessary
  and sufficient conditions},\ }\href
  {https://doi.org/https://doi.org/10.1016/S0375-9601(96)00706-2} {\bibfield
  {journal} {\bibinfo  {journal} {Physics Letters A}\ }\textbf {\bibinfo
  {volume} {223}},\ \bibinfo {pages} {1} (\bibinfo {year} {1996})}\BibitemShut
  {NoStop}%
\bibitem [{\citenamefont {Jha}\ \emph {et~al.}(2021)\citenamefont {Jha},
  \citenamefont {Mukherjee},\ and\ \citenamefont {Bambah}}]{tripartite}%
  \BibitemOpen
  \bibfield  {author} {\bibinfo {author} {\bibfnamefont {A.~K.}\ \bibnamefont
  {Jha}}, \bibinfo {author} {\bibfnamefont {S.}~\bibnamefont {Mukherjee}},\
  and\ \bibinfo {author} {\bibfnamefont {B.~A.}\ \bibnamefont {Bambah}},\
  }\bibfield  {title} {\bibinfo {title} {Tri-partite entanglement in neutrino
  oscillations},\ }\href {https://doi.org/10.1142/S0217732321500565} {\bibfield
   {journal} {\bibinfo  {journal} {Modern Physics Letters A}\ }\textbf
  {\bibinfo {volume} {36}},\ \bibinfo {pages} {2150056} (\bibinfo {year}
  {2021})}\BibitemShut {NoStop}%
\bibitem [{\citenamefont {Hill}\ and\ \citenamefont
  {Wootters}(1997)}]{PhysRevLett.78.5022}%
  \BibitemOpen
  \bibfield  {author} {\bibinfo {author} {\bibfnamefont {S.}~\bibnamefont
  {Hill}}\ and\ \bibinfo {author} {\bibfnamefont {W.~K.}\ \bibnamefont
  {Wootters}},\ }\bibfield  {title} {\bibinfo {title} {Entanglement of a pair
  of quantum bits},\ }\href {https://doi.org/10.1103/PhysRevLett.78.5022}
  {\bibfield  {journal} {\bibinfo  {journal} {Phys. Rev. Lett.}\ }\textbf
  {\bibinfo {volume} {78}},\ \bibinfo {pages} {5022} (\bibinfo {year}
  {1997})}\BibitemShut {NoStop}%
\bibitem [{\citenamefont {Workman}\ \emph {et~al.}()\citenamefont {Workman}
  \emph {et~al.}}]{PDB}%
  \BibitemOpen
  \bibfield  {author} {\bibinfo {author} {\bibfnamefont {R.}~\bibnamefont
  {Workman}} \emph {et~al.} (\bibinfo {collaboration} {Particle Data Group}),\
  }\bibfield  {title} {\bibinfo {title} {{Review of Particle Physics}},\
  }\href@noop {} {\ }\bibinfo {note} {To be published (2022)}\BibitemShut
  {NoStop}%
\end{thebibliography}%

\end{document}